\documentclass[a4paper,fleqn,usenatbib,useAMS]{mnras}

\usepackage{graphicx}	
\usepackage{amsmath}	
\usepackage{amssymb}	
\usepackage{multicol}        
\usepackage{bm}		
\usepackage{pdflscape}	
\usepackage[encapsulated]{CJK}
\usepackage{ucs}
\usepackage[utf8x]{inputenc}
\usepackage{multirow}
\usepackage{booktabs,caption}
\usepackage[flushleft]{threeparttable}
\usepackage[dvipsnames]{xcolor}

\newcommand{\kms}{\,km\,s$^{-1}$} 
 
\usepackage[T1]{fontenc}
\usepackage{ae,aecompl}

\usepackage{newtxtext,newtxmath}

\title[Sab\,19, a PN crossing the Perseus Arm]{
Detailed Studies of IPHAS sources. 
II. 
Sab\,19, a true planetary nebula and its mimic crossing the Perseus Arm.   
}

\author[Guerrero et al.] 
{Guerrero, M.A.$^1$\thanks{E-mail:mar@iaa.es}, Ortiz, R.$^2$, Sabin, L.$^3$, 
Ramos-Larios, G.$^4$, Alfaro, E.J.$^1$\\  
$^1$Instituto de Astrof\'\i sica de Andaluc\'\i a, IAA-CSIC, Glorieta de la Astronom\'\i a, s/n, E-18008, Granada, Spain \\
$^2$Escola de Artes, Ciências e Humanidades, USP, Av.\ Arlindo Bettio 1000, 03828-000 São Paulo, Brazil \\
$^3$Instituto de Astronom\'\i a, UNAM, Apdo.\ Postal 877, Ensenada 22860, B.C., Mexico \\
$^4$Instituto de Astronom\'\i a y Meteorolog\'\i a, CUCEI, Universidad de Guadalajara, Av.\ Vallarta 2602, Arcos Vallarta, 44130 Guadalajara, Mexico 
}

\pubyear{2020}

\begin{document}
\label{firstpage}
\pagerange{\pageref{firstpage}--\pageref{lastpage}}
\maketitle

\begin{abstract}
  
The INT Photometric H$\alpha$ Survey (IPHAS) has provided us with a number of 
new-emission line sources, among which planetary nebulae (PNe) constitute an 
important fraction.  
Here we present a detailed analysis of the IPHAS nebula Sab\,19 
(IPHASX\,J055242.8+262116) based on radio, infrared, and optical 
images and intermediate- and high-dispersion longslit spectra.  
Sab\,19 consists of a roundish 0.10 pc in radius double-shell nebula surrounded by 
a much larger 2.8 pc in radius external shell with a prominent H-shaped filament.  
We confirm the nature of the main nebula as a PN whose sub-solar N/O 
ratio abundances, low ionized mass, peculiar radial velocity, and 
low-mass central star allow us to catalog it as a type III PN.  
Apparently, the progenitor star of Sab\,19 became a PN when crossing 
the Perseus Arm during a brief visit of a few Myr.  
The higher N/O ratio and velocity shift $\simeq$40 \kms\ of the external shell with respect to the main nebula and its large ionized mass suggest that it is not truly associated with Sab\,19, but it is rather dominated by a Str\"omgren zone in the interstellar medium ionized by the PN central star. 

\end{abstract}



\section{Introduction}

The INT Photometric H$\alpha$ Survey (IPHAS: \citealt{Drew2005,Barentsen2014}) 
has mapped the Northern Galactic Plane within the latitude range 
$b \leq |5^{\circ}|$, 
discovering hundreds of new emission-line sources.  
Among those, many can be expected to be planetary nebulae (PNe), and indeed follow up 
spectroscopic observations have unveiled a large sample of new PNe.  
The first release of extended PNe based on the IPHAS catalogue identified 159 
true, likely and possible PNe \citep{Sabin2014}.

We have started a series of detailed analyses of individual IPHAS objects.  
\citet{Sabin2020} and \citet{RG2020} described an evolved bipolar PN and a highly extincted bipolar PN, respectively.  
These two sources at an advanced evolutionary stage and found at large distances and affected by large amounts of extinction can be typically expected among IPHAS PNe.  
Here we have focused our attention in IPHASX\,J055242.8+262116, the source number \#19 in \citet{Sabin2014}'s list that will be referred hereafter as Sab\,19.  
This source, classified originally as a likely PN, is located on the Galactic plane along the Galactic anticenter ($l$=183.0219$^\circ$, $b$=+0.0176$^\circ$) and presents an intriguing triple-shell 
morphology.

We have obtained new images and spectroscopic information for this source and 
combined this information with archival radio and infrared observations.  
Sab\,19 is confirmed to be a true small-size PN surrounded by a much 
larger Str\"omgren zone in the interstellar medium (ISM), which mimics 
a PN halo.  
The article is organised as follows. 
The imaging and spectroscopic observations are listed in \ref{obs}. 
The morpho-kinematics as well as the nebular and stellar properties of Sab\,19 are 
discussed in \ref{results}. 
Finally our discussion on the properties of the PN and its central star and our 
conclusions are presented in \ref{discussion} and \ref{conclusion}, respectively.

\begin{figure*}
\begin{center}
\includegraphics[height=2.7in]{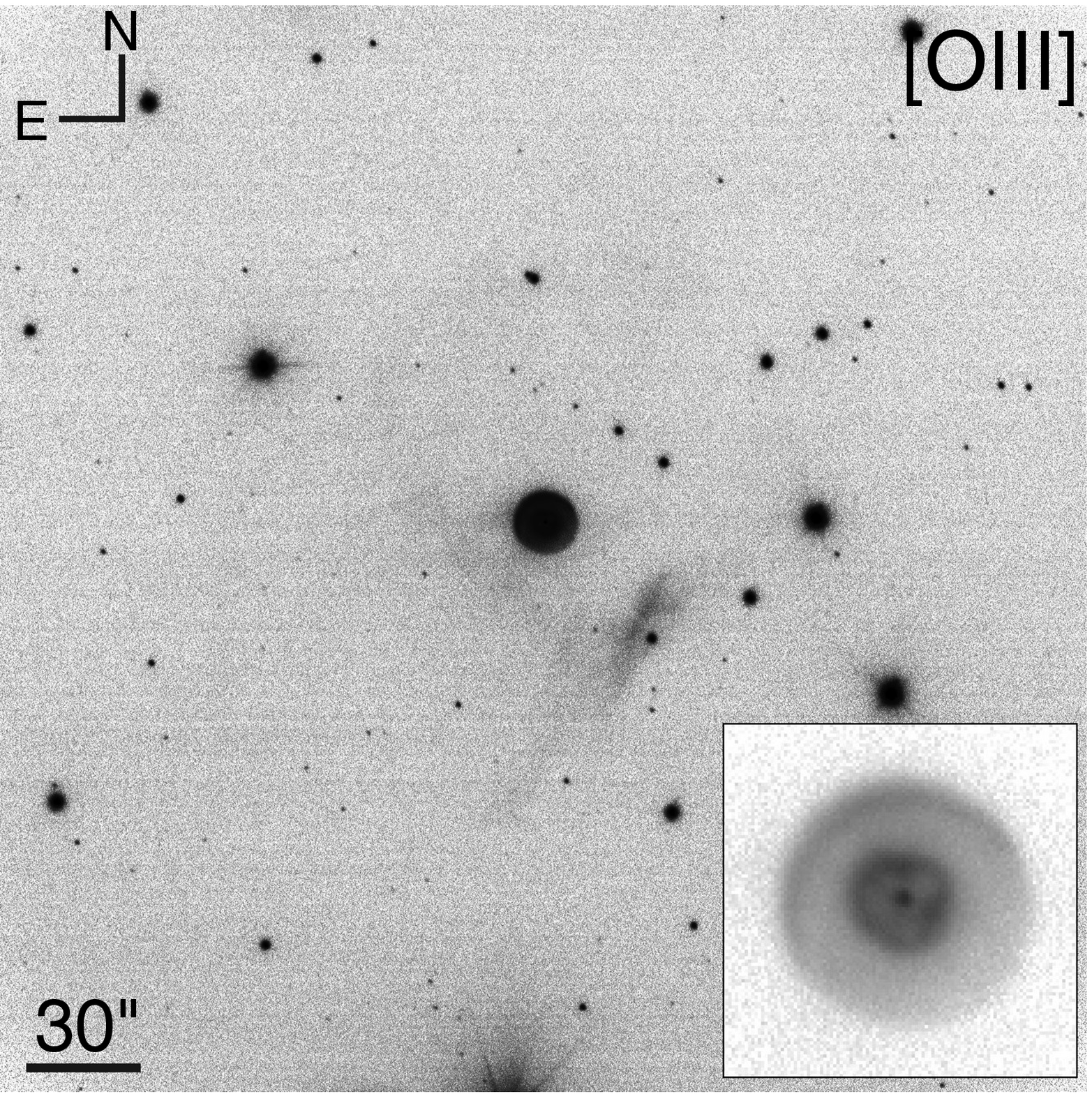}
\hspace{0.075in}
\includegraphics[height=2.7in]{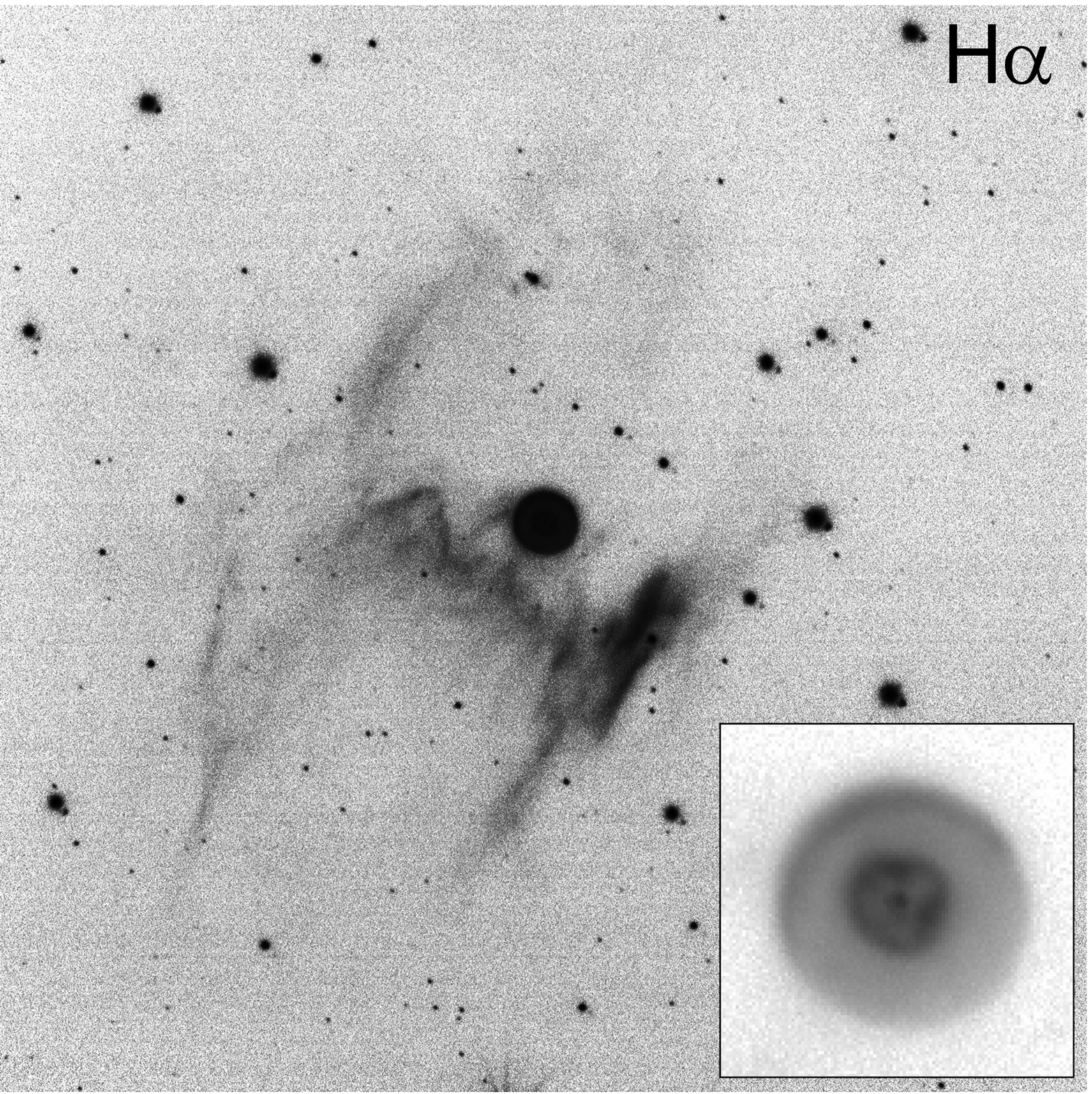}\\
\vspace{0.1in}
\includegraphics[height=2.7in]{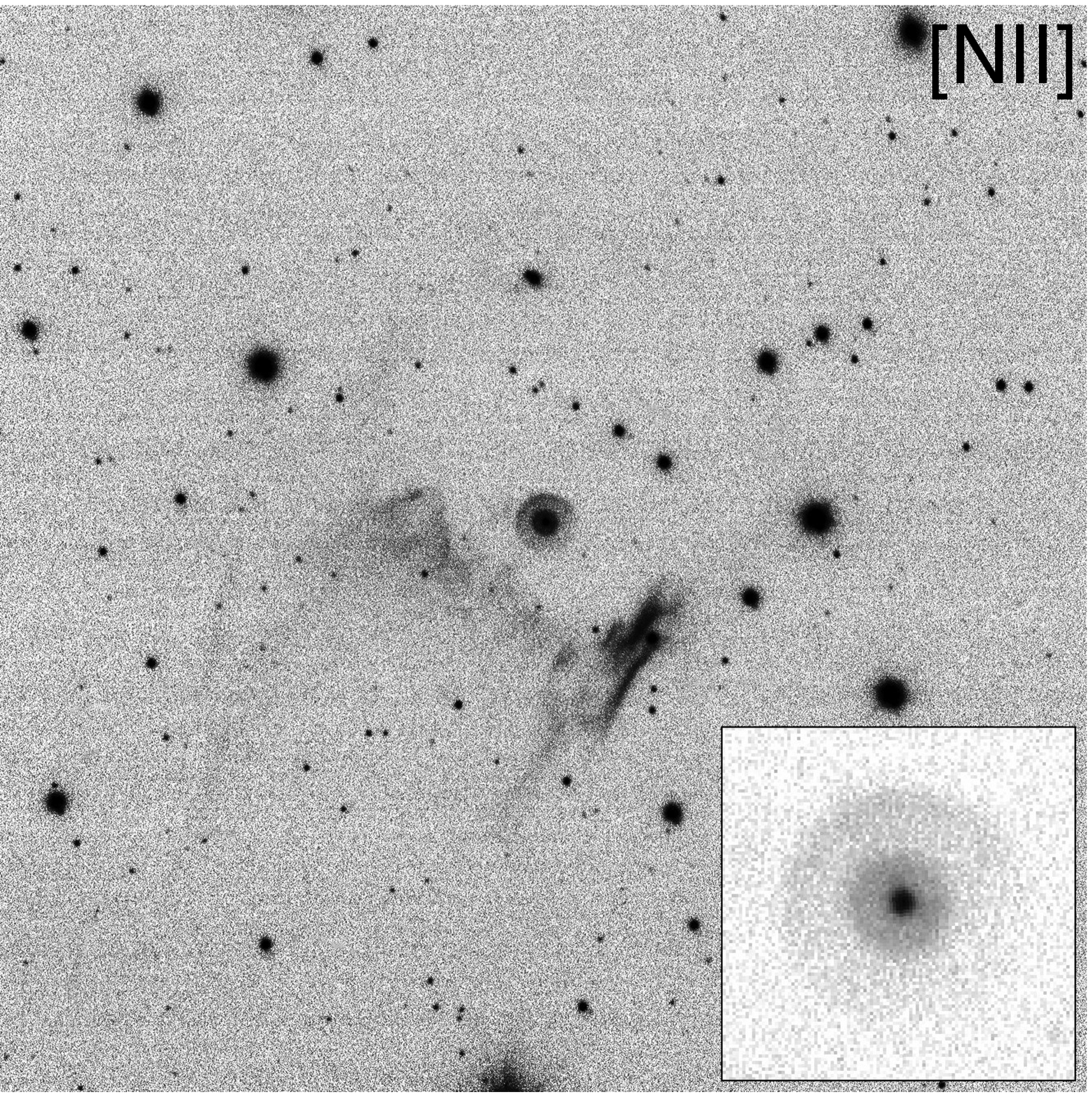}
\hspace{0.075in}
\includegraphics[height=2.7in]{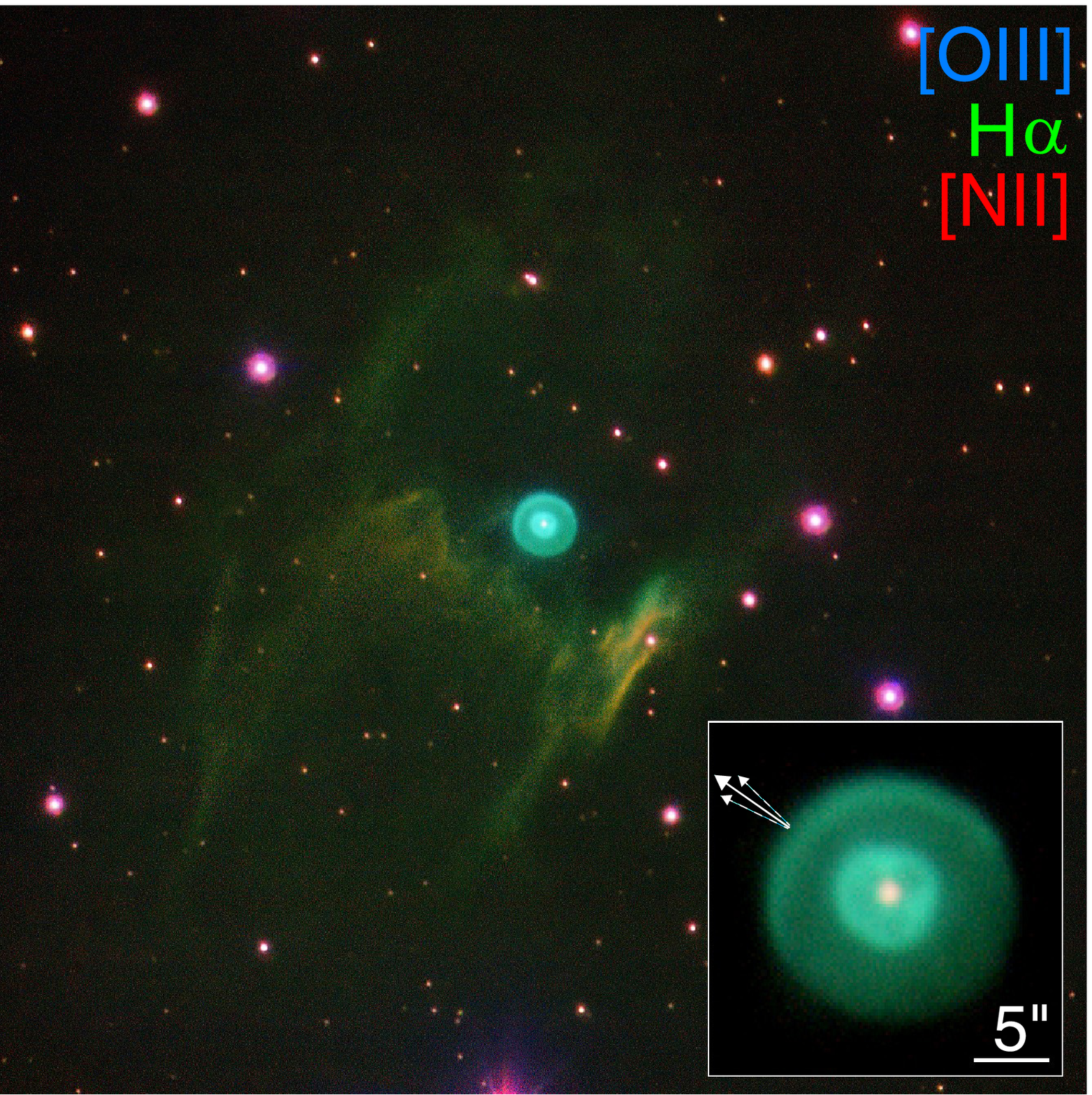}
\vspace{0.02cm}
\caption{
NOT ALFOSC narrowband images of Sab\,19 in the emission lines of [O\,{\sc iii}] $\lambda$5007 \AA, H$\alpha$ and [N\,{\sc ii}] $\lambda$6584 \AA, and colour-composite picture.  
The insets show a zoom-in view of the bright inner nebula. 
The solid white arrow in the inset of the colour picture indicates the direction 
of the motion of Sab\,19 according to {\sc gaia}'s proper motions, with the 
dashed arrows denoting the uncertainty in the direction of the motion. 
}
\label{img:NOT}
\end{center}
\end{figure*}

\section{Observations}\label{obs}

\subsection{Optical Narrowband Imaging}

Narrowband optical images of Sab\,19 in the H$\alpha$, [N\,{\sc ii}] 
$\lambda$6584 \AA, and [O\,{\sc iii}] $\lambda$5007 \AA\ emission lines 
were obtained with the ALhambra Faint Object Spectrograph and Camera 
(ALFOSC) on the 2.5m Nordic Optical Telescope (NOT) at the Roque de los 
Muchachos Observatory (ORM, La Palma, Spain) on January 26, 2020.  
The detector was an E2V 231-42 2k$\times$2k CCD with pixel size 15\,$\mu$m, 
providing a plate scale of 0$\farcs$211 pixel$^{-1}$ and a field of view of 
6$\farcm$3$\times$6$\farcm$3. 
The images were obtained using the Observatorio de Sierra Nevada (OSN) 
H01 H$\alpha$ (\mbox{$\lambda_c = 6565$ \AA}, \mbox{FWHM $= 13$ \AA )} and 
E16 [N\,{\sc ii}] (\mbox{$\lambda_c = 6583$ \AA}, \mbox{FWHM $= 13$ \AA)} 
filters, and the 
NOT \#90 [O\,{\sc iii}] (\mbox{$\lambda_c = 5007$ \AA}, \mbox{FWHM $= 30$ \AA)} filter.

\begin{figure*}
\begin{center}
\includegraphics[width=2.7in]{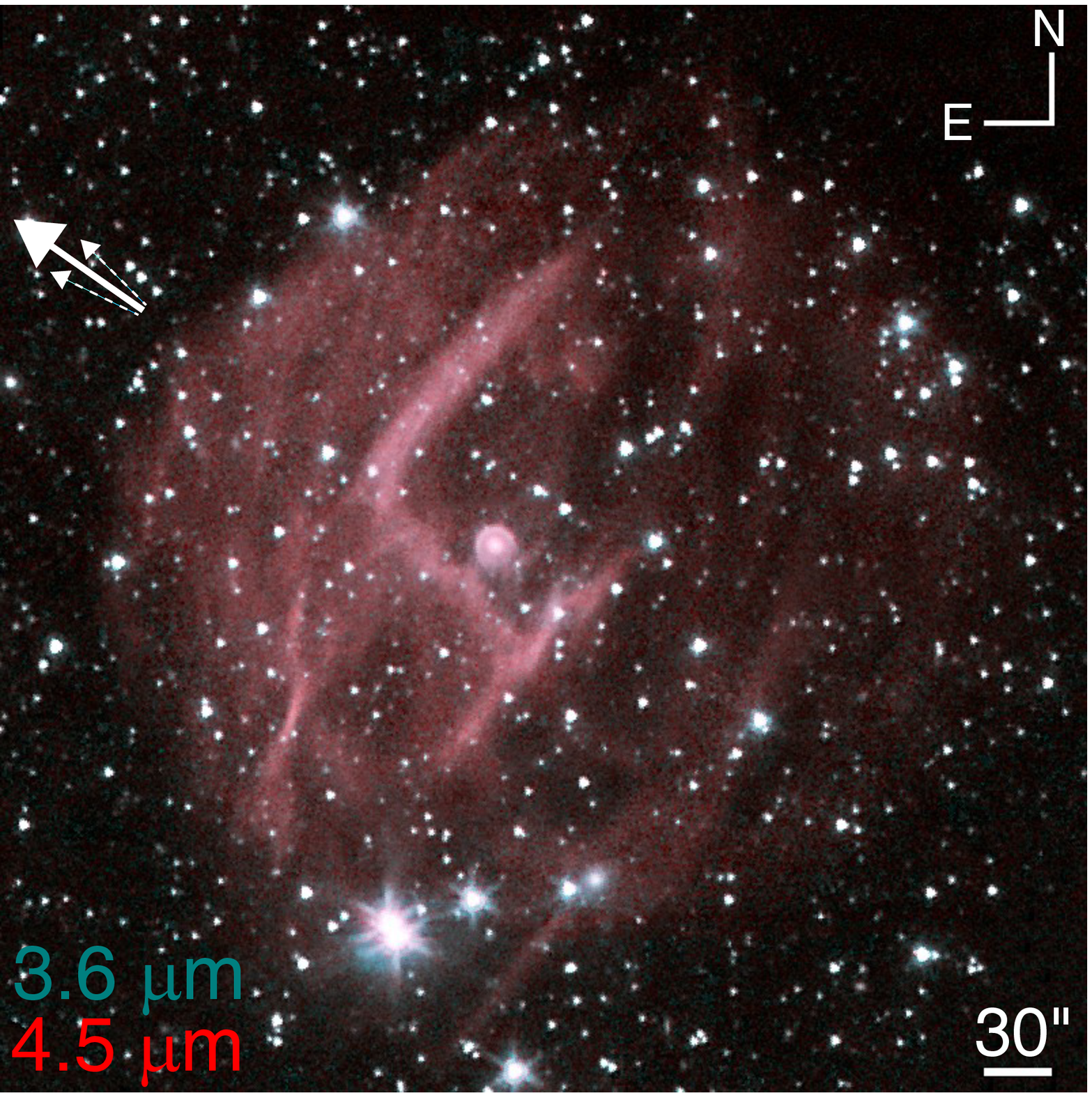} 
\hspace{0.1in}
\includegraphics[width=2.7in]{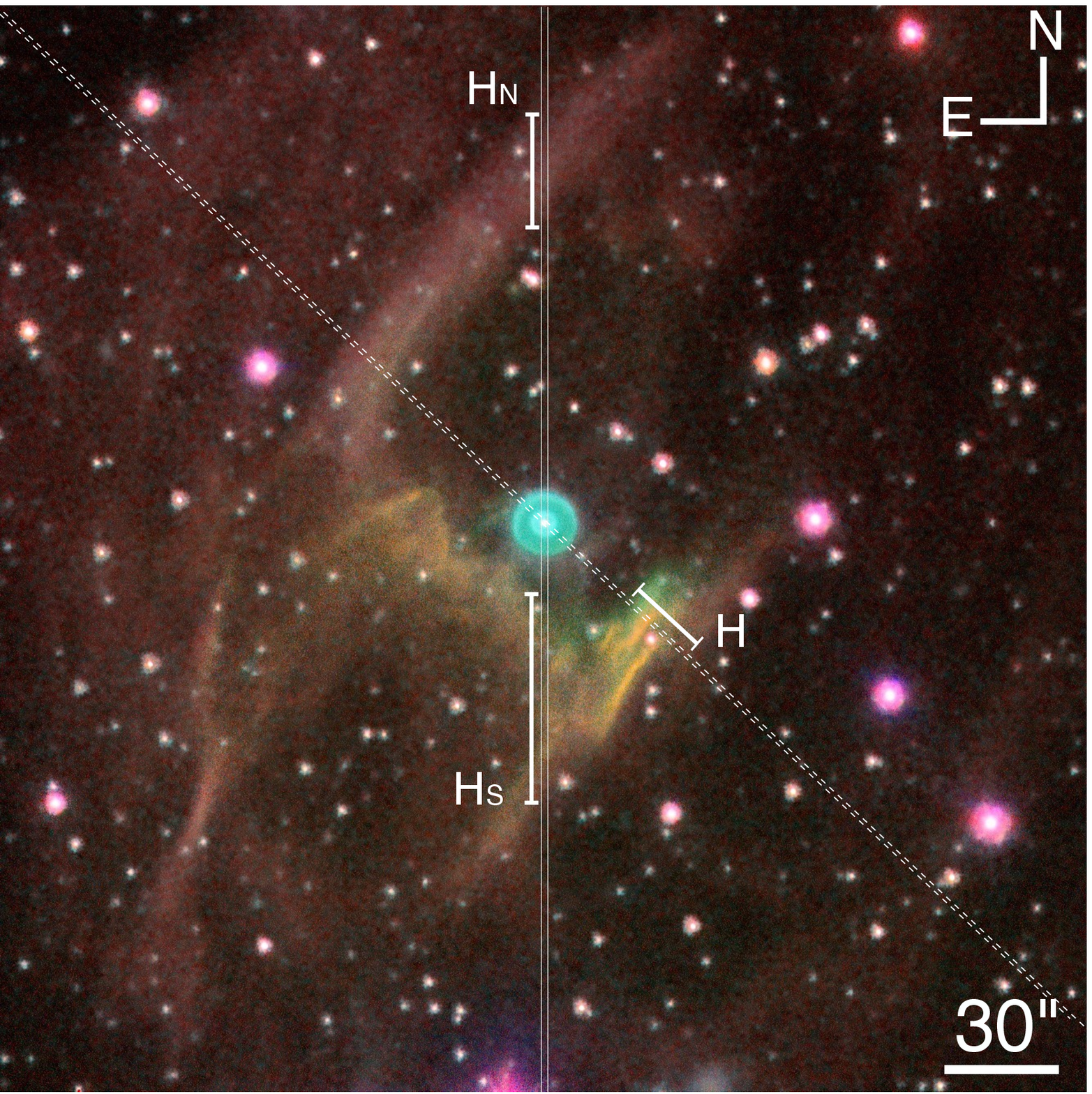} 
\caption{
(left) Colour-composite \emph{Spitzer} IRAC picture of Sab\,19 in the  3.6 (blue-green) and 4.5 (red) $\mu$m bands.  
(right) Colour-composite \emph{Spitzer} IRAC IR and NOT ALFOSC optical picture of Sab\,19.  
The colours in this panel corresponds to those in Figure~\ref{img:NOT} for the optical image and in the left panel for the IR image.  
The arrows in the left panel indicate the motion of Sab\,19 derived from {\sc gaia} observations as described in Figure~\ref{img:NOT}. 
The longslits used for the MES high-dispersion (PA=0$^\circ$) and OSIRIS intermediate-dispersion (PA=47$^\circ$) spectroscopic observations are overlaid, as well as the regions used to extract spectra from the outermost regions.
}
\label{img:Spitzer}
\end{center}
\end{figure*}

Three 600 s exposures were obtained for each filter, with a 
small dithering of a few arcsec between them to improve the 
image quality and remove cosmic rays.  
The observing conditions were excellent with a stable seeing of 1$\farcs$0 for all the images, as inferred from the FWHM of stars in the field of view. 
All the images were reduced using standard {\sc iraf} routines.
After accounting for vignetting caused by the filters, the 
net field of view is $\approx$5\farcm2.

\subsection{Intermediate-dispersion Spectroscopy}

Intermediate resolution spectra were obtained with the 10.4m Gran Telescopio Canarias (GTC) of the ORM with the Optical System for imaging and 
low-Intermediate-Resolution Integrated Spectroscopy \citep[OSIRIS,][]{Cepa2000} 
on January 2, 2018.  
OSIRIS was used with two Marconi 2048$\times$4096 pixels CCD detectors with a 
2$\times$2 binning, leading to a spatial scale of 0$\farcs$254 pixel$^{-1}$.  
The R1000B grism was used, providing a spectral coverage from 3630 \AA\ to 7500 
\AA\ and a dispersion of 2.12 \AA~pixel$^{-1}$.

The observations consisted of four exposures of 450 s and 750 s for a total 
exposure time of 4800 s.  
The slit was set at a position angle (PA) of 47$^{\circ}$ with a length of 7$\farcm$4.  
The slit width of 0$\farcs$8 resulted in a spectral resolution $R$ of 900.   
The data reduction, which includes wavelength calibration with HgAr and Ne lamps and flux calibration with the spectroscopic standard star Feige~110, was performed using standard {\sc iraf} routines.

\subsection{High-dispersion Spectroscopy}

A longslit high dispersion optical spectrum of Sab\,19 was obtained with the
Manchester Echelle Spectrometer (MES, \citealt{Meaburn2003}) mounted on the
2.12m telescope at the Observatorio Astron\'omico Nacional, San Pedro
M\'artir (OAN-SPM, Mexico). 
The observations were obtained on April 17, 2016 with a 2048$\times$2048 
pixels E2V CCD Marconi detector with a pixel size of 13.5 $\mu$m\,pixel$^{-1}$. 
An on-chip 4$\times$4 binning was applied, resulting in a spatial scale 
of 0$\farcs$702 \,pixel$^{-1}$ and a spectral scale 0.11 \AA\,pixel$^{-1}$.

MES provides a slit length of 6$\farcm$5 and the slit width was set to 150 $\mu$m, corresponding to 1$\farcs$9.  
The slit width and spatial scale were suitable for the non optimal weather conditions, with a seeing $\sim$1\farcs8, providing a spectral resolution $\simeq$12 \kms.
The slit was arranged at PA $0^\circ$ and one 
exposure of 1800~s was obtained with an H$\alpha$ filter with 
$\Delta\lambda$ = 90 \AA\ to isolate the 87th echelle order. 
This order also contains the [N\,{\sc ii}] $\lambda\lambda$6548,6584 
\AA\ emission lines. 
A calibration frame of a Th-Ar lamp was obtained immediately 
after the science exposure to perform the wavelength calibration.  
The data were reduced using standard {\sc iraf} routines.

\subsection{Infrared Archival Images}

\subsubsection{Spitzer IRAC images}

Sab\,19 was observed on April 2010 by the \emph{Spitzer} Space Telescope 
under the program GLIMPSE360: Completing the Spitzer Galactic Plane Survey 
(PI: Barbara A. Whitney).  
Images were obtained in the 3.6 and 4.5 $\mu$m channels of the InfraRed Array 
Camera \citep[IRAC,][]{Fazio2004}, a four-channel camera that provides simultaneous 
imaging at 3.6, 4.5, 5.8, and 8 $\mu$m with similar 5\farcm2$\times$5\farcm2 field 
of view (FoV).  
The two short wavelength channels use InSb detector arrays 
consisting of 256$\times$256 pixels and a nearly same pixel scale of 
1\farcs2~pixel$^{-1}$. 
Images were retrieved from the \emph{Spitzer} Heritage Archive.

\subsubsection{WISE imaging}

Wide-field Infrared Survey Explorer Space Telescope 
\citep[\emph{WISE},][]{Wright2010} observations of Sab\,19 in the mid-infrared 
W2 4.6 $\mu$m, W3 12 $\mu$m, and W4 22 $\mu$m  bands were retrieved from the 
NASA/IPAC Infrared Science Archive (IRSA). 
The angular resolutions of these observations were 6\farcs4, 
6\farcs5, and 12\farcs0, respectively.
The cryogenically-cooled telescope uses 1024$\times$1024 detector arrays 
of HgCdTe and Si:As with plate scale of 2\farcs75 pixel$^{-1}$.

\section{Results}\label{results}

\subsection{Morphology}

The optical images in Figure~\ref{img:NOT} reveal an [O~{\sc iii}] and H$\alpha$ 
bright round main nebula with a double-shell morphology and an H-shaped feature 
of diffuse emission brighter in [N~{\sc ii}] and H$\alpha$.  
The main nebula consists of a 6\farcs4$\times$6\farcs0 roundish inner shell 
and a concentric 16\farcs4$\times$15\farcs6 outer shell.  
The outer shell has a notable brightness enhancement towards the North-Northeast 
direction with its apex along PA$\approx$25$^\circ$, whereas the nebular emission 
along the opposite direction fades smoothly. 
The morphology of this outer shell is reminiscent of a bow-shock as caused by the 
motion of the nebula through the ISM.  
This suggestion is reinforced by the offset of the central star (CSPN) 
with respect to the centre of the outer shell by $\sim$0\farcs6 along 
the direction of the apex of the bow-shock.

\begin{figure*}
\begin{center}
\includegraphics[bb=50 150 550 718 ,width=0.90\linewidth]{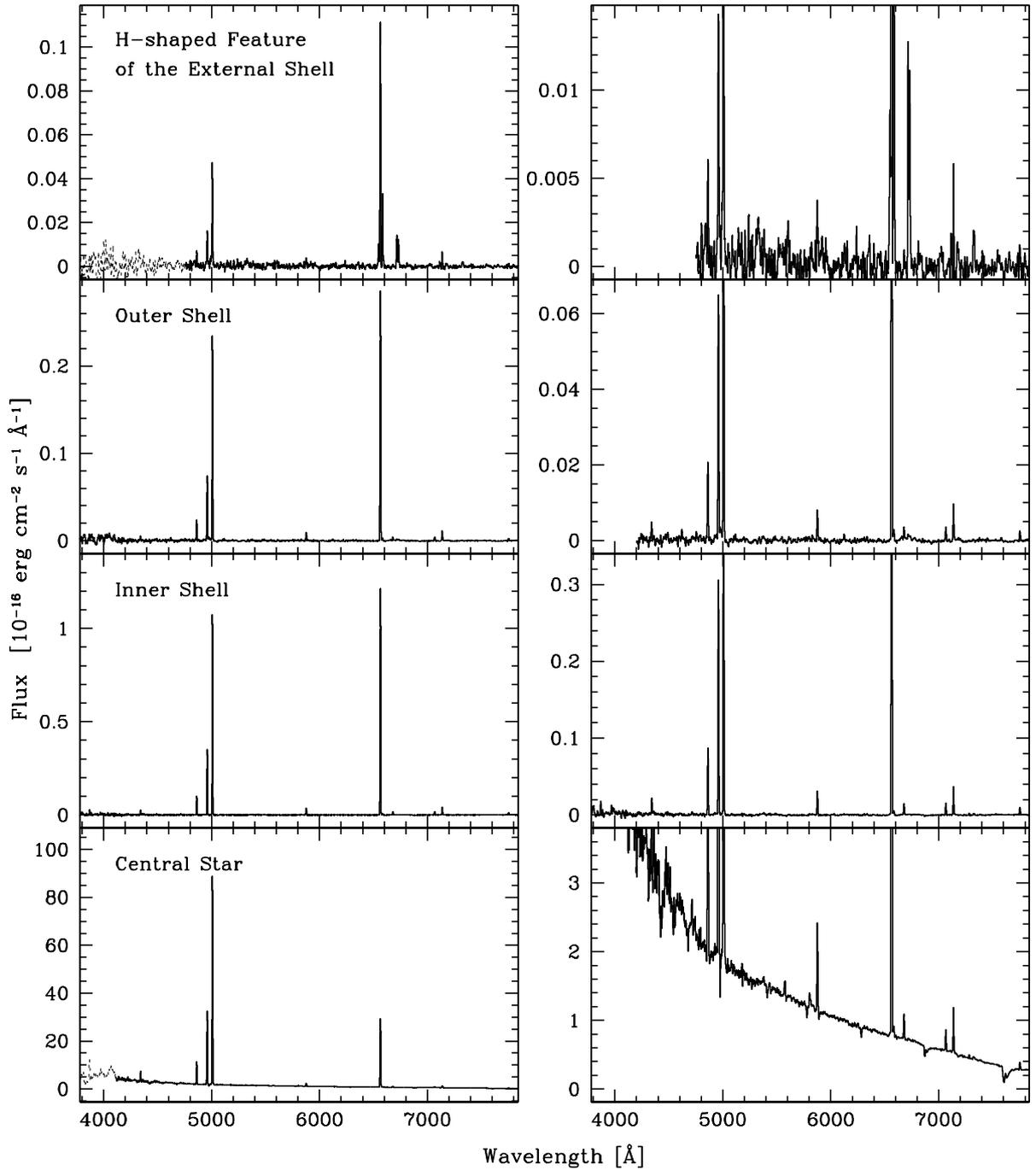}
\caption{
GTC OSIRIS one-dimensional spectra of the H-shaped feature of the external shell, the outer and inner shells, and the central star of Sab\,19 (see the position of the longslit overlaid on Fig.~\ref{img:Spitzer}-right).  
Two different intensity scales are used to show both the bright (left) and faint (rigth) emission lines. 
Spectral regions with low signal-to-noise ratio are shown as dotted lines.  
}
\label{img:OSIRIS_1D}
\end{center}
\end{figure*}

The bright main optical nebula is surrounded by low surface-brightness 
emission with an H-shaped morphology.  
This emission is not centred at the main nebula, but displaced towards 
the South-Southeast with a filamentary and fuzzy appearance.  
The emission is brighter in H$\alpha$, with some bright filaments in [N~{\sc ii}] 
particularly towards the Southwest directon.  
The emission in these different emission lines is clearly stratified, 
with the lower ionization [N~{\sc ii}] emission further away from the 
main nebula.

A comparison with \emph{Spitzer} IRAC images in the available 3.6 and 4.5 $\mu$m 
bands in Figure~\ref{img:Spitzer} reveals that the optical H-shaped feature is not 
only particularly prominent in these IR bands, but it is the brightest feature of 
a larger structure, a roundish shell with a diameter $\approx$7\farcm5.  
This external shell is quite filamentary and, as the outer shell of the main 
nebula, it is brighter towards the Northeast direction and fainter and smoother 
towards the opposite direction.  
This might be indicative of the interaction with the ISM as the 
nebula moves through it, and actually the noticeable displacement 
of the main nebula towards the brightest Northeast rim of this 
external shell lends support to this idea.  
The IR emission in the H-shaped feature also follows the excitation structure 
revealed in the optical emission lines, thus suggesting that the emission from 
this H-shaped feature arises from an atomic, molecular or dusty component at 
lower excitation than the material responsible for the emission of the optical 
[N~{\sc ii}] lines.

\begin{table*}
  \begin{center}
  \caption{Dereddened emission line fluxes with respect to H$\beta$=100.}
  \label{emissionlines}
  \begin{tabular}{lccrrrr}
\hline
Ion & $\lambda$ & $f_\lambda$ & \multicolumn{3}{c}{\underline{~~~~~~~~~~~~~~~~~~PN components~~~~~~~~~~~~~~~~~~}} &  
\multicolumn{1}{c}{External shell} \\
    & (\AA)     &             & 
\multicolumn{1}{c}{CSPN} & 
\multicolumn{1}{c}{Inner Shell} & 
\multicolumn{1}{c}{Outer Shell}  & \\
\hline
$[$Ne\,{\sc iii}$]$ & 3869 &   0.274 & 68$\pm$7~~~   & 57$\pm$7~~~   & $\dots$~~~~   & $\dots$~~~~  \\
H$\gamma$           & 4340 &   0.143 & 36.7$\pm$3.4  & 40.7$\pm$3.7  &  47$\pm$7~~~  & $\dots$~~~~  \\
$[$O\,{\sc iii}$]$  & 4363 &   0.136 & 9.2$\pm$1.2   & $\dots$~~~~   & $\dots$~~~~   & $\dots$~~~~  \\
H$\beta$            & 4861 &   0     &  100$\pm$7~~  & 100$\pm$6~~~  & 100$\pm$10~   & 100$\pm$15~  \\
$[$O\,{\sc iii}$]$  & 4959 & --0.025 &  335$\pm$18~  & 317$\pm$17~   & 278$\pm$24~   & 221$\pm$29~  \\
$[$O\,{\sc iii}$]$  & 5007 & --0.037 & 1000$\pm$40~  & 950$\pm$40~   & 840$\pm$60~   & 570$\pm$60~  \\
C\,{\sc iv}         & 5806 & --0.217 & 2.3$\pm$0.3   & $\dots$~~~~   & $\dots$~~~~   & $\dots$~~~~  \\
He{\sc i}           & 5876 & --0.231 &  15.1$\pm$1.3 & 14.3$\pm$1.2  & $\dots$~~~~   & $\dots$~~~~  \\
$[$N\,{\sc ii}$]$   & 6548 & --0.350 & 0.28$\pm$0.06 & 0.54$\pm$0.09 & 0.52$\pm$0.14 & 26.5$\pm$2.4 \\
H$\alpha$           & 6563 & --0.352 &  286$\pm$20~  & 286$\pm$20~   & 286$\pm$31~   & 286$\pm$40~  \\
$[$N\,{\sc ii}$]$   & 6583 & --0.355 & 1.4$\pm$0.2 & 1.7$\pm$0.2     & 3.4$\pm$0.6   &  84$\pm$14~  \\
He\,{\sc i}         & 6678 & --0.370 & 3.5$\pm$0.4 & 3.4$\pm$0.4     & 3.4$\pm$0.6   & $\dots$~~~~  \\
$[$S\,{\sc ii}$]$   & 6716 & --0.376 & $\dots$~~~~   & $\dots$~~~~   & $\dots$~~~~   & 34.4$\pm$3.1 \\
$[$S\,{\sc ii}$]$   & 6731 & --0.378 & $\dots$~~~~   & $\dots$~~~~   & $\dots$~~~~   & 29.9$\pm$2.7 \\
He\,{\sc i}         & 7065 & --0.426 & 3.3$\pm$0.4   & 3.2$\pm$0.4   & 3.2$\pm$0.5   & $\dots$~~~~  \\
$[$Ar\,{\sc iii}$]$ & 7136 & --0.435 & 7.4$\pm$0.7~  & 7.5$\pm$0.7   & 8.2$\pm$1.2   & 13.7$\pm$2.8 \\
$[$Ar\,{\sc iv}$]$  & 7236 & --0.448 & $\dots$~~~~   & 0.16$\pm$0.04 & $\dots$~~~~   & $\dots$~~~~  \\
He\,{\sc i}         & 7281 & --0.454 & 0.56$\pm$0.09 & 0.46$\pm$0.08 & $\dots$~~~~   & $\dots$~~~~  \\
$[$O\,{\sc ii}$]$   & 7320 & --0.459 & 0.47$\pm$0.08 & 0.37$\pm$0.06 & $\dots$~~~~   & 3.1$\pm$0.7  \\
$[$O\,{\sc ii}$]$   & 7330 & --0.460 & 0.35$\pm$0.06 & 0.32$\pm$0.06 & $\dots$~~~~   & 3.1$\pm$0.7  \\
$[$Ar\,{\sc iii}$]$ & 7751 & --0.513 & 1.3$\pm$0.2   & 1.4$\pm$0.2   & 1.6$\pm$0.3   & $\dots$~~~~  \\
    \hline
log\,F(H$\beta$)   & & & $-$13.8~~~ & $-$14.2~~~ & $-$14.7~~~ & $-$15.3~~~ \\
c(H$\beta$)        &  &    & 1.91$\pm$0.06 & 1.87$\pm$0.06 & 1.81$\pm$0.09 & 2.23$\pm$0.14 \\
T$_e$($[$O\,{\sc iii}$]$) (K)&  &    & 11200$\pm$700 & $\dots$~~~~ & $\dots$~~~~ & $\dots$~~~~ \\
N$_e$($[$S\,{\sc ii}$]$) (cm$^{-3}$) & & & $\dots$~~~~ & $\dots$~~~~ & $\dots$~~~~ & 300$^{+300}_{-200}$ \\
    \hline
  \end{tabular}
 \label{neb}
  \end{center}
\end{table*}

The parallax of Sab\,19 is 0.35$\pm$0.13 mas \citep{Gaia2018}.   
Adopting the bayesian inference method of \citet{Bailer2018}, it implies a distance of 2.6$^{+1.3}_{-0.7}$ kpc.  
Accordingly, the main nebula of Sab\,19 has a radius of 0.10$^{+0.05}_{-0.03}$ pc, whereas its external shell has a radius of 2.8$^{+1.4}_{-0.8}$ pc.

\subsection{Physical Conditions and Chemical Abundances}

The GTC OSIRIS longslit spectra along PA 47$^\circ$ have been used to extract 
one-dimensional spectra of the different structural components of Sab\,19, 
namely the CSPN, the inner and outer shells of the main nebula, and the 
external shell.  
The spectrum of the latter corresponds to the only region detected, the bright filament of the H-shaped feature $\simeq$70$^{''}$ Southwest of the main nebula (marked by the label H in Figure~\ref{img:Spitzer}-right).
The one-dimensional spectra presented in Figure~\ref{img:OSIRIS_1D} 
show generally a small number of emission lines, with very faint or even 
absent emission lines of low ionization species in the main nebula (e.g., 
[S~{\sc ii}]) that become brighter in the external shell.  
We notice that the He~{\sc ii} $\lambda$4686 emission line is not detected 
throughout the nebula, neither in the higher excitation main nebula nor in 
the lower ionization external shell.

\begin{table}
\begin{center}
\caption{Ionic and Elemental Abundances of Sab\,19.}
\label{abundances}
\begin{tabular}{lrr}
\hline
Ratio & Main Nebula & External Nebula \\
\hline
He$^+$/H$^+$    &      0.096$\pm$0.008           & $\dots$~~~~~~~~~ \\
O$^+$/H$^+$     & (6.8$\pm$0.9)$\times$10$^{-6}$ & (1.1$\pm$0.3)$\times$10$^{-4}$ \\
O$^{++}$/H$^+$  & (2.9$\pm$0.1)$\times$10$^{-4}$ & (2.1$\pm$0.3)$\times$10$^{-4}$ \\
N$^+$/H$^+$     & (3.2$\pm$0.5)$\times$10$^{-7}$ & (1.6$\pm$0.3)$\times$10$^{-5}$ \\
S$^+$/H$^+$     &         $\dots$~~~~~~~~~       & (1.6$\pm$0.3)$\times$10$^{-6}$ \\
Ar$^{++}$/H$^+$ & (5.8$\pm$0.7)$\times$10$^{-7}$ & (1.3$\pm$0.4)$\times$10$^{-6}$ \\
\hline
He/H & $\geq$0.096~~~~~~~~~ & $\dots$~~~~~~~~~ \\
O/H  & 3.0$\times$10$^{-4}$~~~~ & 3.3$\times$10$^{-4}$~~~~ \\
N/O  & 0.05$\pm$0.01~~~~~ & 0.15$\pm$0.05~~~~~ \\
\hline
\end{tabular}
\end{center}
\end{table}

The intensity of the emission lines in these spectra relative to H$\beta$ 
and their 1-$\sigma$ uncertainties was determined and dereddened using the extinction law by \citet{Fitzpatrick2007} for $R_V$=3.1 described by the coefficients $f_\lambda$ listed in Table~\ref{neb} 
and the logarithmic extinction coefficient $c$(H$\beta$) also listed in Table~\ref{neb} derived from the observed H$\alpha$ to H$\beta$ line ratio adopting case B recombination.  
The 1-$\sigma$ uncertainties of these logarithmic extinction coefficient imply they have a similar value through the different nebular components, with marginal evidence for a higher extinction in the external shell.

These relative intensities were analyzed using the nebular analysis tool {\sc anneb} \citep{Olguin2011} based on the {\sc iraf} nebular package \citep{Shaw1995}. 
Unfortunately, the spectra display very few temperature or density diagnostic emission lines; 
only a temperature $T_e\simeq$11200~K can be determined from the relative intensities of the auroral-to-nebular [O~{\sc iii}] optical emission lines in the spectrum extracted at the CSPN, and a density $N_e\simeq$300~cm$^{-3}$ from the relative intensities of the [S~{\sc ii}] doublet in the spectrum of the brightest feature of the external shell.

The determination of the ionic and elemental abundances listed in 
Table~\ref{abundances} is thus affected by the small number of 
emission lines in the spectra, but also by the uncertain determination 
of the physical conditions.  
For the main nebula, we have assumed a density of 4000 cm$^{-3}$ as derived from 
radio observations (see Section~\ref{sect:mass}) and adopted a temperature of 
11200 K, whereas for the external shell we have adopted a density of 300 
cm$^{-3}$ and assumed a typical value of 10000 K for the temperature.  
Besides H$^+$ and He$^+$, the main species in the main nebula is O$^{++}$, 
which is almost 90 times more abundant than O$^+$.  
Yet the He~{\sc ii} $\lambda$4686 emission line is not detected, indicating the 
lack of species of high excitation including O$^{3+}$ and above.  
The nebula has thus a peculiar ionization balance, with most of the oxygen atoms as O$^{++}$ ions.  
The He and O abundances are typical of Type~III and IV PNe, with sub-solar N/O ratios \citep[(N/O)$_\odot$=0.25$\pm$0.04,][]{Aetal2009}.  
Although the N/O ratio relies on the determination of the O$^+$ ionic abundances based on the [O~{\sc ii}] $\lambda\lambda$7320,7330 doublet, which is more sensitive to $T_e$ than the determination of the N$^+$ ionic abundances, a solar value for this ratio would require an electronic temperature $\lesssim$8000 K, which seems unlikely for the observed nebular excitation.  
On the other hand, the N/O ratio of the external shell is higher and marginally
consistent with the solar value.

\subsection{Kinematics}
\label{sect:kin}

\begin{figure}
\begin{center}
\includegraphics[bb=90 20 400 600,width=0.70\linewidth]{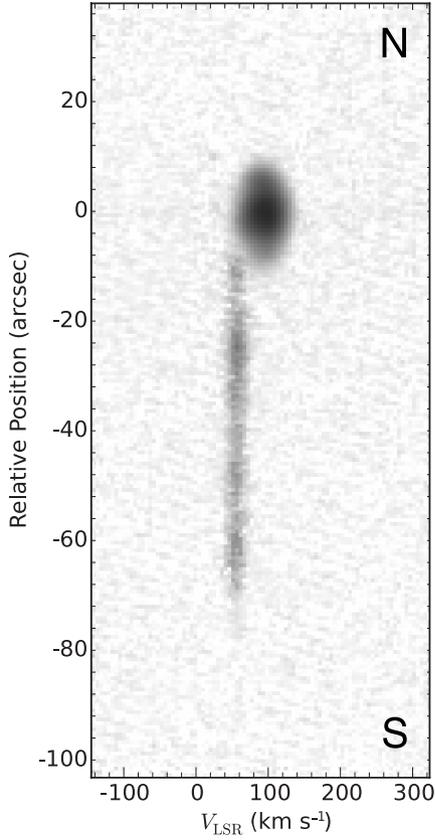}
\caption{
MES H$\alpha$ position-velocity map of Sab\,19 displaying the kinematics of the 
main nebula (lenticular feature at $V_{\rm LSR}\simeq+91$ \kms) and H-shaped 
outer shell (linear feature at $V_{\rm LSR}\simeq+51$ \kms).  
}
\label{img:MES_PV}
\end{center}
\end{figure}

The MES echelle data provide kinematic information of the main nebular shell and the bright Southern region and Norhtern filament of the H-shaped feature intersected by the slit with PA=0$^\circ$ (regions H$_{\rm S}$ and H$_{\rm N}$ in Figure~\ref{img:Spitzer}-right, respectively).
The position-velocity (PV) map of the H$\alpha$ line shown in Figure~\ref{img:MES_PV} reveals a lenticular shape for the main nebula of Sab\,19 with brighter emission in the innermost region associated with its inner shell.  
Meanwhile, the line profile of the H$\alpha$ emission associated with the regions of the H-shaped feature to the South of the main nebula (the Northern filamente is much fainter) is narrower than that of the main nebula and is basically consistent with a unique velocity, i.e., it does not show any velocity structure along the spatial extent $\sim$60$^{\prime\prime}$ of the line.

\begin{figure}
\begin{center}
\includegraphics[bb=130 150 450 718,width=0.80\linewidth]{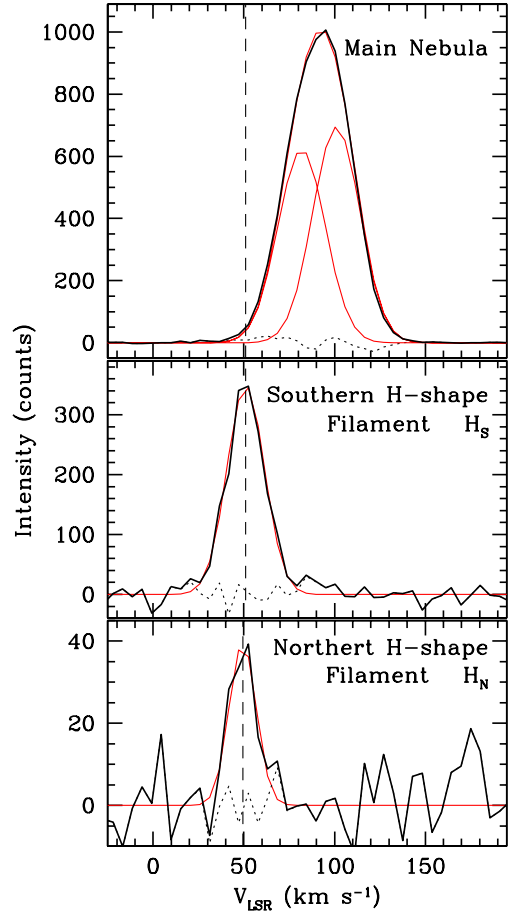}
\caption{
H$\alpha$ line profiles of the main nebular shell (top) and Southern (middle) and Northern (bottom) filaments of the H-shaped feature.  
Gaussian fits to the line profiles are shown as solid red lines and the residuals of the fits as dotted black lines. 
The radial velocity of the H-shaped feature is marked by a vertical dashed line.  
The vertical dashed line shown in the top panel corresponds to the radial velocity of the brighter Southern filament of the H-shape feature. 
}
\label{img:MES_1D}
\end{center}
\end{figure}

The H$\alpha$ line profile extracted from the main shell 
(Fig.~\ref{img:MES_1D}-top) is not resolved, but it is 
broader than the instrumental spectral resolution.  
The line can be fitted with two Gaussian components with radial velocities in the Local Standard of Rest (LSR) $+81.8\pm2.1$ \kms\ and $+101.0\pm2.1$ \kms, implying a LSR radial velocity $V_{\rm LSR} = +91.4\pm1.3$ \kms\ and an expansion velocity $V_{\rm exp} \simeq 9.6\pm3.0$ \kms.  
Adopting this expansion velocity for the outer shell, its 0.10$^{+0.05}_{-0.03}$ pc radius implies a kinematic age 10400$^{+5000}_{-3000}$ yr.

The line profile of the H$\alpha$ emission associated with the H-shaped feature is shown in the middle and bottom panels of Figure~\ref{img:MES_1D}.  
As described above, it is narrow and it can indeed be fitted with a single Gaussian component with radial velocity $V_{\rm LSR} = +51.1\pm1.5$ \kms\ for the Southern region H$_{\rm S}$ and $+49.4\pm3.2$ \kms\ for the Northern region H$_{\rm N}$.
As illustrated in Figures~\ref{img:MES_PV} and \ref{img:MES_1D}, there is a remarkable $\sim$40 \kms\ shift between the radial velocities of the H-shaped feature and the main nebula of Sab\,19.

\begin{figure*}
\begin{center}
\includegraphics[width=0.8\linewidth]{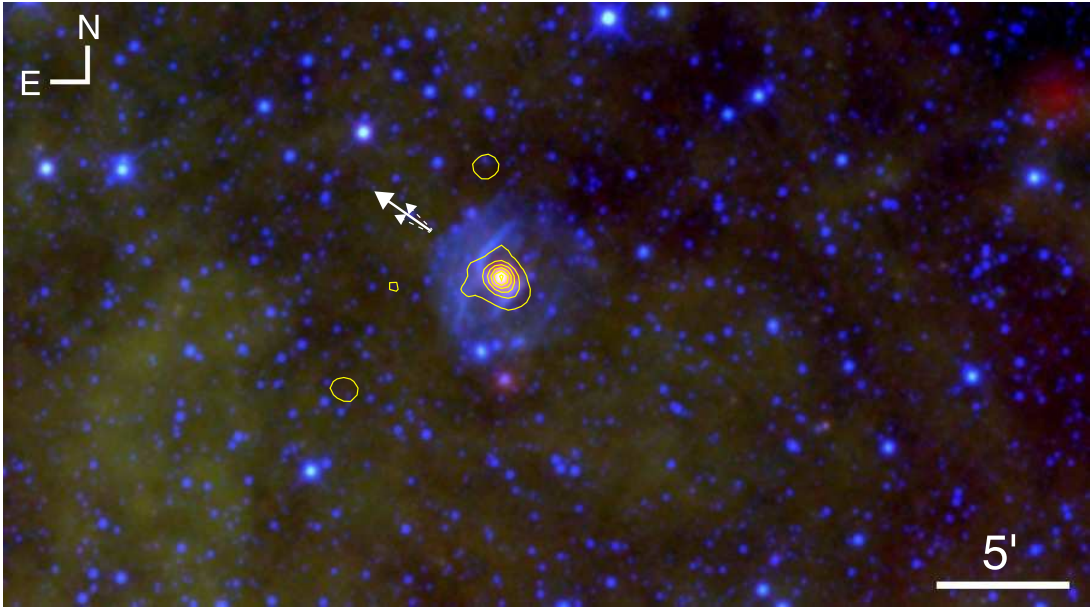} 
\caption{
Colour-composite \emph{WISE} picture of Sab\,19 in the W2 4.6 $\mu$m (blue), 
W3 12 $\mu$m (green) and W4 22 $\mu$m (red) bands overlaid with yellow NVSS 
radio emission contours at 3, 10, 20, 40 and 50 sigmas 
over the background.
The arrows indicate the motion of Sab\,19 derived from {\sc gaia} 
observations as described in Figure~\ref{img:NOT}. 
}
\label{img:wise}
\end{center}
\end{figure*}

\subsection{Nebular Masses}
\label{sect:mass}

The ionized mass $M_{\rm ion}$ of a PN can be estimated from its total H$\beta$ 
flux using, for instance, the relationship 
\begin{equation}
  M_{\rm ion} = 11.06 \times F({\rm H}\beta )\, d^2\, t^{0.88}\, N_e^{-1},
  \label{eq:mass}  
\end{equation}
\noindent where $M_{\rm ion}$ is given in solar masses, 
$F({\rm H}\beta )$ is the extinction-corrected H$\beta$ flux in units of 
$10^{-11}$ erg~cm$^{-2}$~s$^{-1}$, 
$d$ is the distance in kpc, 
$t$ is the electron temperature in units of 10000 K, and 
$N_e$ is the electron density \citep{Pottasch1983}.

The H$\beta$ flux can be derived from the measured H$\alpha$ flux and 
the extinction derived from the nebular spectra $c$(H$\beta$) of 1.9, 
as listed in Table~\ref{emissionlines}.  
The H$\alpha$ flux from the main nebula and the external shell have been derived from the NOT ALFOSC H$\alpha$ image, computing the star-subtracted photon count rate within apertures encompassing the main nebular shell and the external shell (after excising the main nebular shell) and using the GTC OSIRIS spectra to flux calibrate them by comparing the 
image count rate within the nebular area covered by the GTC OSIRIS slit and 
the H$\alpha$ flux derived from this spectrum.  
The observed flux of the main and external nebulae are $1.6 \times 10^{-12}$ 
and $3.6 \times 10^{-12}$ erg cm$^{-2}$ s$^{-1}$, respectively. 
The extinction-corrected intrinsic H$\alpha$ fluxes would be 
$2.9 \times 10^{-11}$ erg cm$^{-2}$ s$^{-1}$ for the main nebula and 
$6.4 \times 10^{-11}$ erg cm$^{-2}$ s$^{-1}$ for the external shell.  
The corresponding intrinsic H$\beta$ fluxes would be 
$1.0 \times 10^{-11}$ erg cm$^{-2}$ s$^{-1}$ for the main nebula and 
$2.2 \times 10^{-11}$ erg cm$^{-2}$ s$^{-1}$ for the external shell\footnote{  
Accounting for the uncertainties in image count rate, cross-calibration of the NOT images with the OSIRIS spectra, and extinction, the total uncertainties for these intrinsic H$\beta$ fluxes are $<$10\% for the main nebula and $<$15\% for the external shell. 
}. 
Therefore, the external shell intrinsic fluxes in these emission 
lines is approximately twice that of the main nebula, although its average surface brightness is $\approx$100 times lower.  
We note that the H$\beta$ intrinsic flux can also be derived from the radio 
flux at 5~GHz using the relationship 
\begin{equation}
\frac{{\rm F}_{5\;{\rm GHz}}}{\rm F(H\beta)} = 2.82 \times 10^9 \, t^{0.53} \, \Big( 1 + \frac{n({\rm He}^+)}{n({\rm H}^+)} \Big)
\end{equation}
\noindent given in \citet{Pottasch1983}.  
The radio flux at this frequency of 44$\pm$4 mJy \citep{GregoryTaylor1986} implies an H$\beta$ flux of $(1.3\pm0.1) \times 10^{-11}$ erg cm$^{-2}$ s$^{-1}$, in general agreement with the previous estimate for the main nebula.

The density of the bright filament of the H-shaped feature of the external shell SW of the main nebula has been estimated to be $\simeq$300 cm$^{-3}$ (Table~\ref{emissionlines}), which is certainly an upper limit to its average density.
If this density is assumed to be the same for all ionized material in this shell, then Eq.~\ref{eq:mass} implies it has an ionized mass $\simeq$0.5 $M_\odot$, that would be a lower limit for the ionized mass according to Eq.~\ref{eq:mass}.
On the other hand, the density of the main nebula is unknown.  
Since its H$\beta$ flux is known, we can use the relationship for the rms density
\begin{equation}
  N_e \; \epsilon^\frac{1}{2} = 
  2.74 \times 10^4 \; \Big( \frac{{\rm F}({\rm H}\beta) \; t^{0.88}}{\theta^3 \; d} \Big) ^\frac{1}{2},
  \label{eq:rms}
\end{equation}
\noindent where 
$\epsilon$ is the filling factor and 
$\theta$ is the angular radius in arcsec \citep{Pottasch1983}.  
Then, for the angular size and H$\beta$ flux of the main nebula, an 
rms density of 800\,$\epsilon^{-\frac{1}{2}}$ cm$^{-3}$ is derived.  
Therefore, according to Eq.~\ref{eq:mass}, the ionized mass of the main nebula 
would be 0.10\,$\epsilon^{\frac{1}{2}}$ $M_\odot$. 

The densities of the main nebula and external shell can also be derived 
from their radio emission.  
Sab\,19 has been detected by several radio surveys carried out at different 
frequencies. 
The NVSS survey \citep{Condon1998} detected it as NVSS\,J055242+262109 
with an integral flux at 1.4 GHz of $45.5\pm2.1$ mJy.  
The source is reported to have an angular size
46$^{\prime\prime}\times28^{\prime\prime}$, 
even though the NVSS images have a 45$^{\prime\prime}$ 
FWHM resolution.  
The {\it Radio Patrol of the Northern Milky-Way} \citep{GregoryTaylor1986}
detected an unresolved 5 GHz source at $17''$ from the central star
of Sab\,19, showing a flux of $44\pm4$ mJy. 
More recently, the AMI Galactic Plane Survey \citep[AMIGPS,][]{Perrott2015} 
detected a 15.7 GHz counterpart with an (integrated) intensity of 33.0 mJy. 
The poor spatial resolution of this survey ($3'$) did not allow an
estimation of its apparent size, and thus it was classified as a 
point source.

Assuming that the radio source associated with Sab\,19 has the same dimensions 
at 1.4, 5, and 15.7 GHz, we can estimate the spectral index $\alpha$, considering
$F(\nu ) \propto {\nu}^{\alpha}$. 
The spectral index between 1.4 and 5 GHz is $-0.027$ and that between 
5 and 15.7 GHz is $\alpha = -0.25$. 
Therefore, at 1.4 GHz the radio continuum is near the turnover point between 
the optically thin and thick regimes, and the optical depth equals unity at 
this frequency \citep{Pottasch1983}:
\begin{equation}
  {\tau}_{\nu}=\int {\kappa}_{\nu}dS = 8.24 \times 10^{-2} T_e^{-1.35}
  {\nu}^{-2.1}\times \int N_pN_e dS = 1, 
\end{equation}
\noindent where 
${\tau}_{\nu}$ is the optical depth at the frequency $\nu$,
${\kappa}_{\nu}$ is the absorption coefficient, 
$T_e$ is the electron temperature (K), and 
$N_p$ and $N_e$ are the proton and electron density (in cm$^{-3}$), respectively, integrated along the line of sight $S$ (in pc). 
Assuming that ${\tau}_{\nu}$ equals the unity at 1.4 GHz, $T_e=10000$~K
and $N_p=N_e$, we obtain the emission measure ($EM$):
\begin{equation}
EM = \int N_e^2dS = 6.18 \times 10^6,
\label{EM}
\end{equation}
\noindent 
where {\it EM} is given in cm$^{-6}$ parsec$^{-1}$.

An inspection of the radio emission in Figure~\ref{img:wise} reveals that the 3$\sigma$ 
contour levels of the emission at 1.4 GHz spreads over an area about $72'' \times 54''$.  
The linear size of this emission, 0.91 pc by 0.68 pc, would correspond to electron densities 2600--3000 cm$^{-3}$.  
If we assume instead that this radio emission uniquely arises from the main nebula 
of Sab\,19\footnote{
As suggested by the coincidence of the H$\beta$ flux derived from these radio observations and that of the main nebula derived from optical observations.}, with a diameter of $16''$ or $0.20$ pc at 2.6 kpc, then the emission 
measure in Eq.~\ref{EM} implies an average electron density $N_e\simeq5600$~cm$^{-3}$.  
Alternatively, if the emission were uniquely associated with the external shell, with a diameter of 7\farcm5 or 5.7 pc at 2.6 kpc, it would imply an average electron density $N_e\simeq1000$~cm$^{-3}$, which is unphysical because it is notably larger than the value derived from the [S~{\sc ii}] line ratio.  
Most likely, the emission measure in Eq.~\ref{EM} is a combination 
of those of the main nebula and external shell.  
The density of the main nebula would be in the range from 2600 to 5600~cm$^{-3}$, 
which is consistent with the estimate of the rms density for values of the filling 
factor $\epsilon$ in the range 0.02--0.10 that would result in ionized masses in 
the range 0.016--0.035 $M_\odot$.

\subsection{Properties of the Central Star}
\label{centralstar}

The excitation of the main nebula of Sab\,19 is quite intriguing and may shed 
some light on the properties of its central star.  
The He~{\sc ii} $\lambda$4686 emission line is not detected with an 
upper limit at 3$\sigma$ of 1.8\% the intensity of the H$\beta$ line.  
The He~{\sc ii} $\lambda$4686 to He~{\sc i} $\lambda$5867 line ratio $\leq$0.12 
thus indicates an effective temperature $\leq$60,000~K, regardless of the optical 
thickness of the nebula \citep{GV2000}.  
The Stoy and Zanstra H~{\sc i} temperatures $\sim$57300$\pm$3400~K and 
$\sim$53000$\pm$10000~K, respectively, derived from the intensity of the 
[O~{\sc iii}] $\lambda$5007 using \citet{KJ1991}'s prescriptions are 
consistent with the above upper limit.  
The stellar spectrum shows the broad spectral feature of C~{\sc iv} $\lambda\lambda$5801,5812 (Fig.~\ref{img:OSIRIS_1D}), but neither 
the C~{\sc iii} $\lambda$5696 nor the O~{\sc vi} $\lambda$5290 
features.  
The range of effective temperatures proposed above for the central star of Sab\,19 
is consistent with these WR features \citep{AN2003}.

\begin{figure}
\begin{center}
\includegraphics[width=8.4 cm]{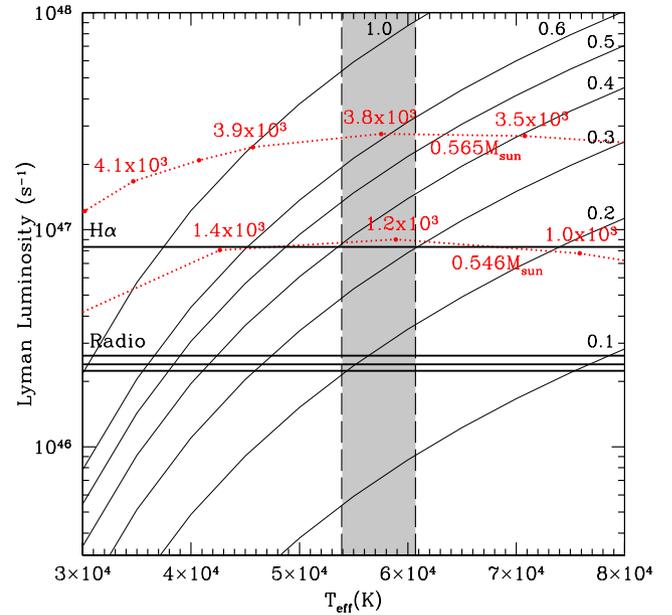}
\caption{
Lyman luminosity of CSPNe as a function of the stellar temperature assuming a blackbody 
model. 
Each curve represents a distinct radius (in $R_{\odot}$).
The thick horizontal lines assign the Lyman luminosity obtained from the H$\alpha$ 
image and from 1.4, 5.0, and 15.7 GHz radio data. 
The red dotted lines represent two evolutionary tracks calculated by 
\citet{Schoenberner1983}, and the quantities along them represent the 
stellar luminosity.
The gray area delimited by the two vertical dashed lines mark the lower and 
upper limits for the stellar temperature, obtained from the spectrum.
}
\label{qpn}
\end{center}
\end{figure}

The Lyman luminosity of the central star ($L_{\rm Ly}$) can be used to 
constrain the stellar radius and luminosity.  
The nebular recombination rate can be obtained from the luminosity in the H$\alpha$ 
line by comparing the total recombination rate with the bound-free transitions down 
to the $n = 2$ level \citep{Osterbrock2006}:
\begin{equation}
  \int_{\nu _o}^{\infty} \frac{L_{\nu}}{h\nu}d\nu =
  \frac{L(H\alpha)}{h\nu _{{\rm H}\alpha}}
  \left(\frac{\alpha_{\rm B}(T)}{\alpha_{{\rm H}\alpha}(T)}\right), 
\end{equation}
\noindent where 
$\alpha_{B,{\rm H}\alpha} (T)$ are the {\it case B} and the {\it n=2} recombination 
coefficients, and 
${\nu}_o$ is the minimum ionization frequency of the hydrogen. 
Adopting the values of $\alpha_{B}=2.59 \times 10^{-13}$ and
$\alpha_{{\rm H}\alpha}=7.69 \times 10^{-14}$ cm$^3$ s$^{-1}$ 
for a temperature of 10000~K \citep{Osterbrock2006}, the intrinsic 
H$\alpha$ luminosities in the main nebula and external shell 
of $2.4 \times 10^{34}$ erg~s$^{-1}$ and $5.2 \times 10^{34}$ erg~s$^{-1}$, 
respectively, derived from the extinction-corrected fluxes 
given in the section above, imply a total of $8.4 \times 10^{46}$
recombinations s$^{-1}$.

At frequencies where the nebula is optically thick to the stellar
Lyman continuum the rate of
incoming ionizing photons over the entire nebula ($L_{\rm Ly}$) can
be related to the free-free emission at the frequency $\nu$ by the
following relationship \citep{Rubin1968}:
\begin{equation}
L_{\rm Ly} = \frac{5.59\times10^{48}}{1+f_i\langle {\rm He^+ / (H^+ + He^+)}
\rangle}\left( \frac{\nu}{5\,{\rm GHz}} \right)^{0.1} T_\mathrm{e}^{-0.45}
F_{\!\nu} d^2,
\end{equation}
\noindent where 
$d$ is the distance in kpc, 
$F_{\!\nu}$ is the total intensity in Jansky, and
$f_i\langle {\rm He^+ / (H^+ + He^+)}$ is the fraction of He-recombination photons 
energetic enough to ionize the H. 
Since the He~{\sc ii} lines are absent in the optical spectrum, we can assume 
the latter to be null.
Assuming $T_\mathrm{e}=10000$~K, we obtain at a distance of 2.6 kpc Lyman 
recombination numbers of 2.4$\times$10$^{46}$, 2.6$\times$10$^{46}$, and
2.2$\times$10$^{46}$ s$^{-1}$ at 1.4, 5 GHz and 15.7 GHz, respectively.

We note that the radio-derived recombinations are about 3.5 times lower than 
that derived using the H$\alpha$ line. 
Several factors might be responsible for this discrepancy. 
For example, the recombination rate obtained from the H$\alpha$ image is strongly 
dependent on the interstellar extinction, whereas the radio value, which is not 
affected by extinction, assumes that Sab\,19 is matter-bounded. 
If the latter hypothesis is incorrect, then $L_{\rm Ly}$ should be considered 
as a lower limit.

\begin{figure}
\begin{center}
\includegraphics[width=8 cm]{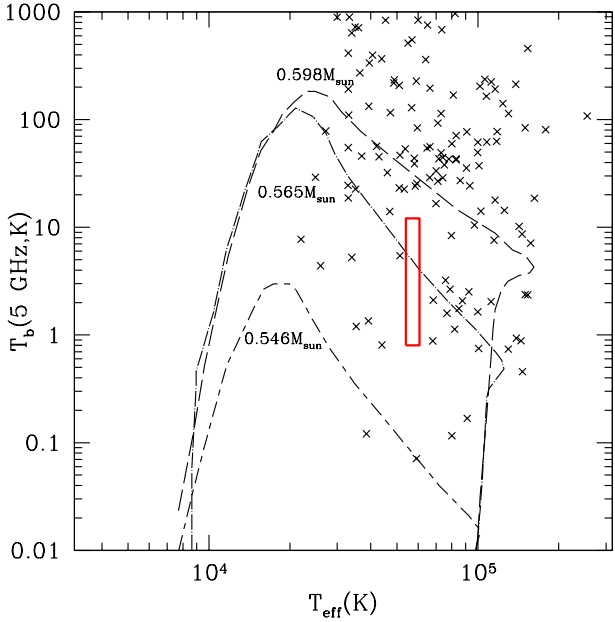}
\caption{
Evolutionary tracks of CSPNe by \citet{ZhangKwok1993}, based on the theoretical models by \citet{Schoenberner1983}.
CSPNe evolve from the left to the right. 
The box marks the lower and upper limits of the temperature of the CSPN ($T_{\rm eff}$) and the brightness temperature of the PN ($T_b$) at 5 GHz, whereas the crosses correspond to the location in this plot of a sample of PNe with measured $T_b$ extracted from \citet{ZhangKwok1993}.
}
\label{Tb}
\end{center}
\end{figure}

These observed $L_{Ly}$ are compared to those computed assuming that the 
CSPN of Sab\,19 emitted like a blackbody for various radii and temperatures 
in Figure~\ref{qpn}.  
The gray area in the plot delimits the upper and lower limits for 
the effective temperature. 
The plot shows also the Lyman luminosity of evolving CSPNe of 0.546 $M_\odot$ and 
0.565 $M_\odot$ calculated from the stellar temperatures and luminosities of the
theoretical tracks by \citet{Schoenberner1983}.
According to this figure, the observed $L_{\rm Ly}$ is compatible with a
very low-mass CSPN, with $M_{\rm CSPN} \le 0.546 M_{\odot}$.
The theoretical tracks indicate that $L_{\rm CSPN} \le 1.2 \times 10^3 L_{\odot}$, 
and the stellar radius is $R_{\rm CSPN}/R_{\odot} = 0.3 - 0.4$, as obtained from 
the H$\alpha$ flux, or $\sim 0.2$, as obtained from the radio data.

\citet{ZhangKwok1993} obtained various distance-independent parameters of PNe 
and analised how they vary as the nebula evolves. 
Their results were based on the theoretical tracks of the CSPN calculated by
\citet{Schoenberner1983}, but they noted that the evolutionary tracks of the 
lowest-mass CSPNe had to be sped up in order to match the observations, which 
is in line with the latest evolutionary models of \citet{Bertolami2016}.   
Figure \ref{Tb} illustrates the evolutionary tracks of a 0.546, 0.565, and 0.598 
$M_{\odot}$ CSPNe by \citet{ZhangKwok1993}, corrected for the ``speed up'' effect. 
During a rapid evolutionary phase the brightness
temperature ($T_b$) of the nebula reaches its maximum, from which it
slowly decreases as the PN evolves. The maximum $T_b$ at 5 GHz
reached by a 0.598, 0.565 and 0.546$M_{\odot}$ nucleus is 200, 120 and
3K, approximately. 
This diagram allows the determination of the mass of the CSPN from two quantities: the nebular brightness temperature and the stellar effective temperature.

The average brightness temperature of Sab19 can be calculated
using  equation
\begin{equation}
  T_b= 0.28 \times \frac{F_{\rm 5~GHz}}{\theta^2}.
\end{equation}
\noindent 
Assuming the dimensions of the 3$\sigma$ contour levels at 1.4 GHz shown in 
Figure~\ref{img:wise} ($72'' \times 54''$), we obtain a lower limit for $T_b$ 
of 0.80~K, whereas an upper limit of 12.1 K is obtained assuming that all 
radiation emitted at 5 GHz originates in the main nebula only. 
When plotted in Figure~\ref{Tb}, these results indicate 
$M_{\rm CSPN} \simeq 0.56 M_{\odot}$. 
Considering the uncertainties, this result agrees with the previous estimates 
obtained using the H$\alpha$ and radio nebular fluxes shown in Figure~\ref{qpn}.

\section{Discussion}\label{discussion}

\subsection{The True and the Mimic}

The analysis of the observations presented in previous sections has revealed 
a multi-component structure for Sab\,19 consisting of a main double-shell 
nebula and a larger and fainter external shell with a prominent H-shaped 
feature.  
Their distinct properties suggest a distinct nature for each 
morphological component.

The main nebula of Sab\,19 has a typical double-shell PN morphology.  
Its spectrum and the physical conditions and chemical abundances derived 
from it are also typical of PNe.  
It can be concluded that the main nebula of Sab\,19 is a true PN.  
The nebula has sub-solar N/O abundances ratio, which is suggestive of a type III 
or IV PN \citep{MK1994}, although it must be noticed that uncertainties in the 
values of O/H and N/O are large.  
The nebula has a small ionized mass, in the range from 0.016--0.035 $M_\odot$, 
and its central star is relatively cold with a low mass, 0.546--0.565 $M_\odot$.
Apparently, the main nebula of Sab\,19 is a type III PN descendant 
from a low-mass progenitor.

The external shell might then be interpreted as a halo resulting from 
an enhanced mass-loss episode associated with a thermal pulse in the last 
phases of the AGB, but this does not seem to be the case.  
First, the total ionized mass of the external shell, which has been 
estimated to be $\gtrsim$0.5 $M_\odot$, is in sharp contrast with 
the low mass of its progenitor star.  
Most notably, the large radial velocity shift between this external shell 
and the main nebula, $\simeq$40 km~s$^{-1}$, casts serious doubts on its 
nature as a PN halo, as true haloes of PNe do not exhibit such large 
velocity shifts with their PNe \citep{GVM1998}.  
Possible mimics of PNe include a long list of sources, as discussed in 
detail by \citet{FP2010}.  
Given the velocity discrepancy between the external shell and the main 
nebula of Sab\,19 \citep[as is also the case of the nebula PHL\,932,][]{F2010} 
and its location inside a much larger mid-IR patchy structure (Fig.~\ref{img:wise}), 
the most likely candidate for the external shell of Sab\,19 among the usual 
suspects for PN mimics is a Str\"omgren zone in the ISM.  
To further reassure this possibility, the relationship between Sab\,19 
and the local ISM is investigated in the next section.

\subsection{Sab\,19 and Its Place in the Galaxy}

At a distance of 2.6$^{+1.3}_{-0.7}$ kpc along galactic longitude 
$\simeq$183$^\circ$ (i.e., mostly at the Galactic anticenter), 
Sab\,19 is located in the outskirts of the Perseus Arm, which 
extends from 1.4 to 2.9 kpc along this direction, as traced by 
very young high-mass stars \citep{Reid2019}.  
This description is consistent with the image in Figure~\ref{img:wise}, which 
shows large-scale patchy emission in the WISE W3 band at 12 $\mu$m revealing 
a complex ISM along the line of sight of Sab\,19.  
Since the mid-IR emission from the external shell of Sab\,19 peaks in the 
\emph{Spitzer} 3.6 $\mu$m and \emph{WISE} W2 4.6 $\mu$m bands, whereas that 
of the ISM around it is brighter in the \emph{WISE} W3 12 $\mu$m band 
(Figs.~\ref{img:Spitzer} and \ref{img:wise}), it can be concluded that the 
mid-IR emission from the external shell of Sab\,19 includes line emission 
or is indicative of warmer dust than that in the ISM.

The positional coincidence of Sab\,19 with the Perseus arm is, however, 
in sharp contrast with their respective radial velocities.  
The LSR velocity at the Galactic anticenter is expected to be null.  
Actually, LSR velocities in the range from $-$20 km~s$^{-1}$ to 
$+$10 km~s$^{-1}$ are measured in the Galactic H~{\sc i} emission 
and giant molecular clouds along this direction 
\citep[see Figure~3 in][]{Reid2019}.  
For instance, WB717 is a CO (J=1$-$0) source detected towards Sab\,19 at 
$V_{\rm LSR}=-0.2$ km s$^{-1}$, whereas other neighbouring CO sources 
detected within Galactic longitude $\pm 1^{\circ}$ from Sab\,19 exhibit 
a range of velocities $-13$ km~s$^{-1} < V_{\rm LSR} <$ +9 km~s$^{-1}$ 
\citep{WB1989}.
The radial velocity of the main nebula of Sab\,19, however, 
is notably different, $V_{\rm LSR}=+91$ km s$^{-1}$.

The discrepancy in radial velocity is also notorious on the tangential 
component of the velocity, i.e., the velocity on the plane of the sky.  
The proper motion components of Sab\,19 measured by {\sc gaia} are 
$pm_{\rm RA}$  = 1.234$\pm$0.205 mas~yr$^{-1}$ and 
$pm_{\rm DEC}$ = 0.848$\pm$0.161 mas~yr$^{-1}$, 
implying a large angle with the Galactic plane.  
On the other hand, the proper motion module of 1.50 mas~yr$^{-1}$ 
corresponds to a linear velocity on the plane of the sky $\simeq$19
\kms\ for the distance given by \citet{Bailer2018}.  
This confirms that the motion of Sab\,19 is dominated by its radial 
velocity and that it does not corrotates with the Perseus Arm, 
reinforcing the idea that Sab\,19 is not actually associated with 
it, but it is crossing it at a relative velocity $\gtrsim$80 \kms.  
Actually, the distance of 2.6 kpc of Sab\,19 would place it in the outer 
border of the Perseus Arm, which it would be "leaving" after a "short visit" 
of a few Myr 
\citep[15 Myr if a thickness of 1.5 kpc is adopted for the Perseus arm 
according to Figure~3 in][]{Reid2019}, but we reckon that the distance 
error bar towards Sab\,19 makes this claim uncertain.

The complete information on the position of Sab\,19 in the phase space has 
been used to analyse its dynamic properties using the {\sc galpy} package 
\citep{Bovy2015}.
The initial values of the orbit are given by the coordinates, distance, 
proper motion and radial velocity of Sab\,19, thought we reckon the large 
uncertainties in distance and proper motions.  
The nebula is then assumed to move under the gravitational potential 
MWPotential2014 included in {\sc galpy}, whose structure and physical 
properties are discussed in detail by \citet{Bovy2015}.  
As for our position and velocity in the Galaxy, the solar motion [--9.4, 12.6, 6.3] 
in km~s$^{-1}$, height over the Galactic plane of $h$ = 16 pc, distance to the 
Galactic centre of 8.15 kpc and circular rotation speed at the Sun’s position of 
$\theta_0$ of 236 km~s$^{-1}$ have been adopted \citep{EAC2006,ECA2006,Reid2019}. 
According to this model and initial conditions, Sab\,19 has a  
highly-elongated orbit, with an apogee at almost 15 kpc from the 
Galactic centre and its perigee at 9 kpc. 
The orbit is contained in the Galactic disk, although it is able to reach a 
maximum height over the Galactic plane $\simeq$700 pc, and the rotational 
velocity is $\leq$250 \kms. 
We note that, for the whole distance and proper motion error intervals, 
the orbit remains confined to the Galactic disk.

\subsection{Just Passing By, but Leaving a Mark}

The notable differences between the radial velocity of the main nebula of 
Sab\,19 and the external shell make very unlikely the latter to be a halo 
ejected in late phases of the stellar evolution.  
Interestingly, the radial velocity of the external shell 
($V_{\rm LSR}\simeq+51$ km~s$^{-1}$) is half way between 
that of the main nebula of Sab\,19 
($V_{\rm LSR}\simeq+91$ km~s$^{-1}$) and that expected 
for the ISM 
($-13$ km~s$^{-1} < V_{\rm LSR} <$ +9 km~s$^{-1}$).  
This seems to imply that the material in this external shell has experienced an interaction with the moving PN.

To investigate into more detail the possible interactions caused by the motion of Sab\,19 through the ISM, the direction of its motion according to the {\sc gaia} proper motions of 
$pm_{\rm RA}$  = 1.234$\pm$0.205 mas~yr$^{-1}$ and 
$pm_{\rm DEC}$ = 0.848$\pm$0.161 mas~yr$^{-1}$ has been overplotted on 
Figures~\ref{img:NOT}, \ref{img:Spitzer}, and \ref{img:wise}.  
The coincidence of the orientation of morphological asymmetries in the 
different shells of Sab\,19 and the direction of its motion on the plane 
of the sky lends support to this interaction: 
the central star is offset towards the SW from the main nebula, which 
shows a bow-shock-like brightness enhancement at the NE direction 
(Fig.~\ref{img:NOT}), the infrared shell is compressed along this direction 
and shows a more diffuse and fainter emission along the opposite SW trailing 
direction (Figs.~\ref{img:Spitzer} and \ref{img:wise}), and the radio 
emission shows a bow-shock-like morphology towards the NE and a smooth decline 
in brightness towards the SW along the trailing direction.

The interactions of the outer shells and haloes of PNe with the surrounding 
ISM have been largely reported and described \citep[e.g.,][]{TK96,WZO2007}.  
Indeed, the location of Sab\,19 in the Perseus Arm and the large-scale 
IR emission around it indicates that the local ISM is relatively dense.  
The external shell of Sab\,19 is not a halo, however, as most (if not all) 
the material in this shell belongs to a Str\"omgren zone in the ISM which 
is being ionized by the CSPN of Sab\,19.  
Still, the H-shaped filaments in this external shell may arise as the result of 
Rayleigh-Taylor (RT) instabilities formed by the interaction of a fast-moving 
ionized shell with a cold, dense, perhaps magnetized ISM. 
The shape of the structures formed depend on many parameters, such as the 
intensity of the magnetic field, the pitch angle between the velocity of 
the PN and the local magnetic field, etc.  
For a fast-moving PN, the shock formed between the PN and the local ISM is 
isothermal, and the ISM magnetic field can contribute to form RT instabilities
just behind the shock wave \citep{SD1997}.  
At any rate, RT structures normally do not penetrate the ionized main PN 
shell, but they are restricted to the external, fragmented external shell 
\citep{SD1997,DS1998}, as is the case of Sab\,19.

It is interesting to note that the module of the velocity vector of Sab\,19 on 
the plane of the sky, $\simeq$19 km~s$^{-1}$, is about 5 times smaller than its 
radial component along the line of sight, $\simeq$91 km~s$^{-1}$.  
We can thus expect that the external shell of Sab\,19 would be much 
thicker along the line of sight than its projection on the plane of 
the sky.  
If we keep in mind that the velocity component of Mira on the plane of the sky is $\simeq$130 \kms, quite similar to the radial velocity of Sab\,19, it could be envisaged the external shell of Sab\,19 as the long tail left behind by Mira \citep{Martin2007}, but seen face on.  
A similar structure also on the plane of the sky might be observed in the PN HFG\,1 \citep{Boumis2009,Chiotellis2016}.

\subsection{The Nature of Sab\,19}

The peculiar velocity of Sab\,19 and the maximum height of its orbit over 
the Galactic Plane can be used to further investigate its past evolution.  
It has been noted that the average peculiar velocity increases among the 
different Peimbert's types of PNe \citep{Peimbert1978}, increasing from 
20$\pm$14 \kms\ for the N- and He-rich type I PNe descending from more 
massive progenitors up to 170$\pm$80 \kms\ for type IV PNe of the halo 
evolving from low-mass progenitors \citep{MD92}.  
The maximum height over the Galactic Plane of Sab\,19 implies that it is not 
a halo type IV PNe, but it can be rather classified as a type III PN.  
The peculiar velocities of type III PNe are indeed typically high, 
$\sim$60 \kms, with many of them showing higher peculiar velocities 
such as Me\,2-2 ($-140$ km~s$^{-1}$), IC\,5217 ($87$ km~s$^{-1}$), 
or K\,3-67 ($-84$ km~s$^{-1}$).

\citet{OM94} observed that AGB stars can also be classified into types similar 
to the Peimbert scheme according to their kinematics. 
The theoretical models of \citet{VW94} and \citet{Bertolami2016} can be used to 
estimate the mass of the precursor ZAMS stars for a set of metallicities. 
The 0.546–0.565 $M_\odot$ mass of the CSPN of Sab\,19 is consistent with the 
0.528-0.652 $M_\odot$ final mass of a solar metallicity (Z=0.02) ZAMS star 
with an initial mass 1.00-1.25 $M_{\odot}$.  
Adopting a lower metallicity of Z=0.001, the final mass of 0.534-0.552 $M_\odot$ 
implies initial masses of ZAMS stars in the range 0.90-1.00 $M_{\odot}$. 
Therefore, according to these theoretical models and the mass of the CSPN 
determined in Sect. \ref{centralstar}, Sab\,19 descends from a nearly 
solar-mass main-sequence star.  
Its maximum rotational velocity of 250 \kms\ makes it very likely 
a member of the thin disk population.

\section{Summary and Conclusion}\label{conclusion}

We have presented new and archival multi-wavelength images and new 
intermediate- and high-dispersion spectroscopic observations of 
Sab\,19, aka IPHASX\,J055242.8+262116.  
These observations reveal that Sab\,19 consists of two morphological components, 
a double-shell main nebula and an external shell dominated by a prominent set of 
H-shaped filaments.  
At a distance of 2.6 kpc, as derived from {\sc Gaia}, the size of these shells is 0.10 and 2.8 pc, respectively.  
The origin of these two components is different.

The double-shell main nebula is a type III PN descending from a low-mass $0.90-1.25 M_\odot$ progenitor star, according to the small nebular ionized mass, the 0.56 M$_\odot$ low-mass CSPN, the peculiar velocity of the nebula, its small expansion velocity, and the nebular sub-solar N/O ratio.  
On the other hand, the higher N/O ratio, large ionized mass and radial velocity shift with the main nebula of the external shell suggested by the present data make it very likely to be a Str\"omgren zone in the ISM ionized by the CSPN of Sab\,19.  
A complete spatio-kinematic study and investigation of the physical conditions in the external shell will certainly help to confirm it.

Sab\,19 is located in the Perseus arm, but its peculiar radial velocity 
implies that it is not associated with it, but it is actually crossing it 
as it moves towards the apogee of its galactic orbit at about 15 kpc from 
the Galactic centre.  
Apparently, the progenitor of Sab\,19 is a low-mass star of the thin disk on 
a very eccentric orbit that happened to become a double-shell PN as it was 
crossing the Perseus Arm.  
The interactions of the PN and its associated Str\"omgren zone 
with the local ISM are quite notorious.

\section*{Acknowledgements}

MAG acknowledges support from the Spanish Government Ministerio de Ciencia, Innovaci\'on y Universidades (MCIU) through grant PGC2018-102184-B-I00, LS from UNAM DGAPA PAPIIT project IN101819, GR-L support from Consejo Nacional de Ciencia y Tecnolog\'\i a (CONACyT) grant 263373 and Programa para el Desarrollo Profesional (PRODEP) Mexico, and EJA from MCIU grant PGC2018-095049-B-C21.  
MAG and EJA are supported by the State Agency for Research of the Spanish MCIU through the “Center of Excellence Severo Ochoa” award for the Instituto de Astrof\'\i sica de Andaluc\'\i a (SEV-2017-0709).

We appreciate Dr.\ V.M.A.\ G\'omez-Gonz\'alez for helpful discussion on the nature and properties of the central star of Sab\,19.

This research has made use of the SIMBAD database operated at CDS (Strasbourg, 
France), the NASA/IPAC Infrared Science Archive, which is funded by the National 
Aeronautics and Space Administration and operated by the California Institute of 
Technology, the NASA's Astrophysics Data System and NRAO VLA Sky Survey.
It makes also use of data obtained as part of the INT Photometric H$\alpha$ Survey 
of the Northern Galactic Plane (IPHAS) carried out at the Isaac Newton Telescope 
(INT), which is operated on the island of La Palma by the Isaac Newton Group in 
the Spanish Observatorio del Roque de los Muchachos of the Instituto de Astrof\'\i 
sica de Canarias. 
All IPHAS data are processed by the Cambridge Astronomical Survey Unit, at the 
Institute of Astronomy in Cambridge. 
IAA ALFOSC guaranteed time

\section*{Data Availability}

All data used in this manuscript are available through public archives, but the 
NOT narrow-band images that can be obtained from the authors after a justified 
requirement.





\end{document}